\documentstyle{report}
\catcode`@=11 
\def\thechapterhead{\arabic{chapter}}
\def\thesectionhead{\thechapterhead.\arabic{section}}

\def\@sect#1#2#3#4#5#6[#7]#8{\ifnum #2>\c@secnumdepth
     \def\@svsec{}\else                                           
     \refstepcounter{#1}\edef\@svsec{\csname the#1head\endcsname\hskip 1em }\fi
     \@tempskipa #5\relax
      \ifdim \@tempskipa>\z@ 
        \begingroup #6\relax
          \@hangfrom{\hskip #3\relax\@svsec}{\interlinepenalty \@M #8\par}
        \endgroup
       \csname #1mark\endcsname{#7}\addcontentsline
         {toc}{#1}{\ifnum #2>\c@secnumdepth \else
                      \protect\numberline{\csname the#1head\endcsname}\fi
                    #7}\else
        \def\@svsechd{#6\hskip #3\@svsec #8\csname #1mark\endcsname
                      {#7}\addcontentsline               
                           {toc}{#1}{\ifnum #2>\c@secnumdepth \else
                             \protect\numberline{\csname the#1head\endcsname}\fi
                       #7}}\fi
     \@xsect{#5}}

\def\appendixname{Appendix}     
\def\appendix{\par
 \setcounter{chapter}{0}
 \setcounter{section}{0}
 \def\@chapapp{\appendixname} 
 \def\thechapter{\Alph{chapter}}
 \def\thechapterhead{\Alph{chapter}}
 \def\thesection{\thechapter.\arabic{section}}
 \def\thesectionhead{\thechapterhead.\arabic{section}}}
\def\@makechapterhead#1{ {\setbox0=\hbox{#1} \parindent 0pt 
 \begin{raggedright} 
 \Large \bf 
 \ifnum \c@secnumdepth >\m@ne \ifdim \wd0=0pt \@chapapp{} \thechapterhead
   \else \thechapterhead \hspace{1ex} \ #1\fi \end{raggedright}\par 
 \fi \nobreak \vskip 2ex }}
\def\@makeschapterhead#1{ { \parindent 0pt \begin{raggedright}
 \Large \bf  #1 \end{raggedright}\par 
 \nobreak \vskip 2ex } }
\def\thebibliography#1{\chapter*{\bibname\@mkboth 
 {\uppercase{\bibname}}{\uppercase{\bibname}}}\list 
 {[\arabic{enumi}]}{\settowidth\labelwidth{[#1]}\leftmargin\labelwidth 
 \advance\leftmargin\labelsep 
 \usecounter{enumi}} 
 \def\newblock{\hskip .11em plus .33em minus -.07em} 
 \sloppy 
 \sfcode`\.=1000\relax} 
\def\bibname{References} 
\def\chapter{\par \vspace{3.5ex} \thispagestyle{plain} \global\@topnum\z@
\@afterindentfalse \secdef\@chapter\@schapter}
\def\section{\@startsection {section}{1}{\z@}{-2.0ex plus -1ex minus
 -.2ex}{1.0ex plus .2ex}{\Large\bf}}
\def\subsection{\@startsection{subsection}{2}{\z@}{-2.0ex plus -1ex minus
 -.2ex}{0.5ex plus .2ex}{\large\bf}}
\def\citenum{\@ifnextchar [{\@tempswatrue\@zcitex}{\@tempswafalse\@zcitex[]}}
\def\@zcitex[#1]#2{\if@filesw\immediate\write\@auxout{\string\citation{#2}}\fi
  \def\@citea{}\@zcite{\@for\@citeb:=#2\do
    {\@citea\def\@citea{,\penalty\@m\ }\@ifundefined
       {b@\@citeb}{{\bf ?}\@warning
       {Citation `\@citeb' on page \thepage \space undefined}}%
\hbox{\csname b@\@citeb\endcsname}}}{#1}}
\def\@zcite#1#2{{#1\if@tempswa , #2\fi}}

\def\citevb{\@ifnextchar [{\@tempswatrue\@citexvb}{\@tempswafalse\@citexvb[]}}
\def\@citexvb[#1]#2{\if@filesw\immediate\write\@auxout{\string\citation{#2}}\fi
  \def\@citea{}\@cite{\@for\@citeb:=#2\do
    {\@citea\def\@citea{--\penalty\@m}\@ifundefined
       {b@\@citeb}{{\bf ? \@citeb}\@warning
       {Citation `\@citeb' on page \thepage \space undefined}}%
\hbox{\csname b@\@citeb\endcsname}}}{#1}}
\newbox\uphookbox
\setbox\uphookbox=\hbox{\vrule width 1.2pt depth 3.0pt height 1.18pt}
\dp\uphookbox=0\p@\def\uphook{\copy\uphookbox}

\def\uphookbracefill#1#2{%
\setbox1=\hbox{$\m@th\displaystyle#1$\hskip-1.2pt}%
\setbox2=\hbox{$\m@th\displaystyle#2$\hskip-1.2pt}%
$\m@th\hskip0.5\wd1\uphook\leaders\vrule\hfill\uphook\hskip0.5\wd2$}
\def\uphookbrace#1#2#3{\mathop
   {\vbox
      {\ialign{##\crcr
         \noalign{\kern3\p@}%
         \uphookbracefill{#1}{#3}\crcr
         \noalign{\kern3\p@\nointerlineskip}%
         $\hfil\displaystyle{#1#2#3}\hfil$\crcr}
      }%
   }%
\limits}
\def\sqr#1#2{{\vcenter{\hrule height.#2pt
\hbox{\vrule width.#2pt height#1pt \kern#1pt
\vrule width.#2pt}
\hrule height.#2pt}}}

\newif\ifseparateabstract\separateabstractfalse
\def\abstract#1{\gdef\@abstract{#1}}
\def\pacs#1{\gdef\@pacs{#1}}
\def\maketitle{\begin{titlepage}%
 \let\footnotesize\small      
 \setcounter{page}{0}%
 \null
 \vfil
 \vskip 30pt                  
 \vbox                                            
  {\begin{center}    
   {\Large\bf \@title \par}        
   \vskip 3em                  
   {                    
     \lineskip .75em
     \begin{tabular}[t]{c}\@author 
     \end{tabular}\par}      
    \vskip 1.5em               
   { \@date \par}        
  \end{center} \par}\nopagebreak 
  \@thanks                                                              
 \setcounter{footnote}{0}       
 \ifseparateabstract 
   \vfil\null                                                
   \end{titlepage}   
   \begin{titlepage}
   \null\vfil
 \else
   \vskip 0pt plus 0.6fil\penalty5000
 \fi
 \vbox
   {\begin{center}                                                  
     {\large\bf Abstract}       
   \end{center}\nopagebreak
   \@abstract    
   \par\vskip4ex \noindent\@pacs} 
 \vfil\null\end{titlepage}
\let\thanks\relax
\gdef\@thanks{}\gdef\@author{}\gdef\@title{}\let\maketitle\relax
\gdef\@abstract{}\gdef\@pacs{}}
\gdef\@thanks{}\gdef\@author{}\gdef\@title{}
\gdef\@abstract{}\gdef\@pacs{}
\catcode`@=12 
\newcommand{\asymx}{\mathop{\sim}}
\newcommand{\asym}[1]{\mathrel{\asymx_{#1}}}

\newcommand{\half}{{\textstyle \frac{1}{2}}}
 
\newdimen\savebaselineskip 
\savebaselineskip=\baselineskip 

\catcode`@=11
\def\lsim{\mathrel{\mathpalette\@versim<}}
\def\gsim{\mathrel{\mathpalette\@versim>}}
\def\@versim#1#2{\vcenter{\offinterlineskip
        \ialign{$\m@th#1\hfil##\hfil$\crcr#2\crcr\sim\crcr } }}
\def\ltgt{\mathrel{\mathpalette\@vergt<}}
\def\gtlt{\mathrel{\mathpalette\@verlt>}}
\def\@vergt#1#2{\vcenter{\offinterlineskip
        \ialign{$\m@th#1\hfil##\hfil$\crcr#2\crcr>\crcr } }}
\def\@verlt#1#2{\vcenter{\offinterlineskip
        \ialign{$\m@th#1\hfil##\hfil$\crcr#2\crcr<\crcr } }}
\catcode`@=12

\newcommand{\natnum}{{\rm I}\mkern-4mu{\rm N}}

\newcommand{\s}{s}

\begin{document}
\title{Applications of the Dotsenko-Fateev Integral in  
       Random-Matrix Models}
\author{
     {\bf Peter~J.~Forrester \& Josef~A.~Zuk\thanks{Postal address: 
     Aeronautical Research Laboratory, Air Operations Division, 
     P.O.~Box~4331, Melbourne 3001, Australia.
     (E-mail~address:~Josef.Zuk@dsto.defence.gov.au)}} \\
     {\em Department of Mathematics, University of Melbourne,
     Parkville, Victoria 3052, Australia}    
       }
\date{}
\pacs{PACS numbers: 05.40.$+$j, 71.23.$-$k, 71.27.$+$a, 24.60.$-$k 
     \hfill January 1996}
\abstract{
The characteristic multi-dimensional integrals that represent physical 
quantities in random-matrix models, when calculated within the 
supersymmetry method, can be related to a class of integrals introduced in
the context of two-dimensional conformal field theories by Dotsenko \&
Fateev. Known results on these Dotsenko-Fateev integrals provide a means
by which to perform explicit calculations (otherwise difficult) 
in random-matrix theory. 
We illustrate this by (i) an evaluation of the mean squared S-matrix elements
for the Gaussian orthogonal ensemble coupled with $M$ external channels, and
(ii) a direct derivation of the asymptotic behaviour of the dynamical
density-density correlator in the limit of large spatial and temporal
separation for the Calogero-Sutherland model which, at certain couplings, 
is known to map onto the parameter-dependent random matrix ensembles   
}
\maketitle 

\chapter{The GOE Problem with External Channels}
\vspace*{-2.0ex}
\section{Background}
Various problems in quantum physics involving classical chaos
or disorder can be modelled by Hamiltonians belonging to one of the 
standard ensembles of Gaussian-distributed random matrices \cite{Meh91} 
coupled to (non-chaotic) external
channels. The random Hamiltonian models the chaotic dynamics of the
internal system --- which may be the compound nucleus in a nuclear
scattering problem \cite{VWZ}, a ballistic electron-microstructure in the 
shape of a classically chaotic cavity \cite{PWZL}, or a disordered
mesoscopic wire exhibiting diffusive electron motion \cite{IWZ}. 
In the example from nuclear physics, the external channels
represent the open decay modes of the compound nucleus; while in the
mesoscopic conductance problems, they enumerate the transverse modes of
the electron wave-function which enter or leave the chaotic system through
attached ideally-conducting leads. 

The central physical observable of relevance in such systems is the
S-matrix, for which the external channels constitute the asymptotic
states. In nuclear-physics applications, the cross-section is related to
the square of S-matrix elements. 
In applications from condensed-matter physics, it is the conductance that
can be obtained by summing the squares of the S-matrix elements
describing transmission. This relationship arises from the Landauer
formula, viz., 
\begin{equation}
g = \sum_{a=1}^{M/2}\sum_{b=M/2+1}^M \{|S_{ab}|^2+|S_{ba}|^2\} \;, 
\label{Landauer}
\end{equation}
where the index $a$ sums over the $M/2$ incoming channels and $b$ sums 
over the $M/2$ outgoing channels. 
Here, $g$ here denotes the dimensionless conductance, and is related to
the physical conductance $G$ by 
\mbox{$g = h/e^2{\cdot}G$}. 
There is, in fact, a close analogy between the statistical behaviour of 
cross-section and conductance for such random-matrix problems. 
For example, there appear
parallel universal characteristics in the two types of system: Ericsson
fluctuations in nuclear cross-sections \cite{BW} and universal conductance
fluctuations in mesoscopic physics \cite{IWZ,LSF}. 
These comparisons are discussed more fully in Ref.~\citenum{HAW}. 

Thus, in order to study the statistical properties of the cross-section or
conductance, one is interested in calculating the ensemble averages of
moments of S-matrix elements over the distributions of the random
Hamiltonian matrices. 

The coupling of the random system with the external channels 
\mbox{$c = 1,2,\ldots,M$}
is fully characterized by the so-called sticking probabilities $T_c$,
which range between zero and unity. The value of $T_c$ can be loosely
viewed as an impedance mismatch for a given channel, which interpolates 
between total instantaneous reflection and perfect transmission for 
values of zero and unity, respectively. 
The $T_c$ are derived as expressions involving combinations of all the
microscopic parameters of the lead-system interface. These quantities are
generally unknown and one does not try to specify them. Rather,
assumptions are made directly about the sticking probabilities, based on
the macroscopic characteristics of the experimental set-up. 
In most cases, the channels are physically equivalent, and so 
it is usual to take all the sticking probabilities to be
equal. Also, in mesoscopic conductance problems, in the absence of
detailed information, one typically sets the value of this common sticking
probability to unity, thereby assuming perfect electron transmission at
the interface. This represents a maximally symmetric situation that is
amenable to exact mathematical analysis. 

With the simplifications stated above having been made, the ensemble
average of a typical squared S-matrix element $|S_{ab}|^2$ is independent 
of the channel indices $a,b$ provided 
\mbox{$a \neq b$}. 
Furthermore, in the case of the Gaussian Orthogonal Ensemble (GOE), 
the presence of 
elastic enhancement (or equivalently --- in mesoscopic physics --- 
coherent back-scattering) ensures that 
\mbox{$\overline{|S_{aa}|^2} =2\overline{|S_{ab}|^2}$} 
for all 
\mbox{$a \neq b$}. 
Application of the supersymmetry method \cite{Efetov,VWZ} to problems with
GOE symmetry in general yields analytical results for 
$\overline{|S_{ab}|^2}$
in terms of superintegrals over a 16-dimensional coset-manifold of 
\mbox{$8\times 8$} 
supermatrices $Q$. The superintegration ranges over eight commuting
(real) degrees of freedom and eight anti-commuting (Grassmann) variables.
These superintegrals can normally be reduced further to 
characteristic triple (ordinary) integrals over one compact parameter
and two non-compact (semi-infinite) ones. 
These parameters can be identified (modulo coordinate transformations)
with the distinct eigenvalues of the supermatrices $Q$. 
For the problem at hand, one obtains the expression given in Eq.~(C.13) of
Ref.~\citenum{PWZLW}, namely
\begin{equation} 
\overline{|S_{ab}|^2} = {\textstyle \frac{1}{4}}I_M \;, 
\label{IM}
\end{equation} 
provided 
\mbox{$a \neq b$}, 
where $I_M$ denotes the integral
\begin{eqnarray}
I_M & = & \half \int_0^1 d\mu \int_1^\infty d\mu_1 \int_1^\infty d\mu_2\, 
     \frac{|\mu_1-\mu_2|}{[(\mu_1-1)\mu_1(\mu_2-1)\mu_2]^{1/2}}{\cdot} 
     \frac{\mu^M}{(\mu_1\mu_2)^{M/2}} 
\nonumber \\
& & \cdot \frac{\mu(1-\mu)}{(\mu_1-\mu)^2(\mu_2-\mu)^2}\left[ 
     \frac{2}{\mu} - \frac{1}{\mu_1} - \frac{1}{\mu_2}\right] \;. 
\label{superint}
\end{eqnarray}
We should note that the change of variables 
\begin{equation} 
\mu\mapsto1-\mu \;, \quad 
     \mu_1\mapsto\mu_1-1 \;, \quad 
     \mu_2\mapsto\mu_2-1 
\label{chvar}
\end{equation} 
has been made relative to Ref.~\citenum{PWZLW}. 

It will be advantageous to make the further change of integration variables
\begin{equation} 
\mu = \frac{1}{t} \;, \quad 
     \mu_1 = \frac{1}{t_1} \;, \quad 
     \mu_2 = \frac{1}{t_2} \;, 
\end{equation} 
so that we can rewrite $I_M$ as 
\begin{equation} 
I_M  =  \int_1^\infty dt \int_0^1 dt_1 \int_0^1 dt_2\, 
     \frac{|t_1-t_2|}{[(1-t_1)(1-t_2)]^{1/2}} {\cdot} 
     \frac{(t_1t_2)^{M/2}}{t^M} 
     {\cdot}\frac{t-1}{(t-t_1)^2(t-t_2)^2} {\cdot} (t-t_1) \;, 
\label{IMT} 
\end{equation} 
where we have made use of the fact that the integrand of $I_M$ is
symmetric in the variables $\mu_1$ and $\mu_2$. 
     
By appealing to the general principles of (i) unitarity of the S-matrix 
and of (ii) 
elastic enhancement in the presence of orthogonal symmetry, one can
argue that the result 
\begin{equation} 
\overline{|S_{ab}|^2} = 1/(M+1)
\label{OneOnM}
\end{equation}
must hold, assuming 
\mbox{$T_c = 1$} 
for all
\mbox{$c = 1,2,\ldots,M$}, 
as we have done \cite{PWZLW}. 
Evaluating the integrals in Eq.~(\ref{superint}) directly
is quite another matter, in spite of the simplicity of the final result. 
Let us also note here that the ensuing expression for the mean dimensionless
conductance $\overline{g}$ under the present assumptions is 
\begin{equation}
\overline{g(M)} \quad=\quad \frac{M^2}{2}\overline{|S_{ab}|^2}
     \quad=\quad \frac{M^2}{8}I_M \;. 
\label{gvalue}
\end{equation}

The integral in Eq.~(\ref{superint}) has been evaluated explicitly in
Appendix~C of Ref.~\citenum{PWZLW} after a lengthy and convoluted
calculation using a method that will not readily generalize upon the
introduction of additional complications. 
One such complication that is of physical interest is the incorporation of a
parameter $\s$ which measures the breaking of orthogonal symmetry due to
the presence of a magnetic field and which drives the transition towards 
behaviour described by the Gaussian Unitary Ensemble. 
If one were to imagine expanding the corresponding result for 
$\overline{|S_{ab}|^2}$ 
in powers or inverse powers of $\s$, in order to study the limiting cases
of small amd large magnetic fields, respectively, then the coeffients in
such an expansion would also be triple integrals of the same general
nature, which one could expect to evaluate analytically as algebraic
expressions in the number of channels $M$. 

Such terms were evaluated using laborious {\it ad hoc} procedures in
Appendix~D of Ref.~\citenum{PWZLW}. 
It would be very useful to have a general systematic method for 
performing these calculations, as well as those for the higher dimensional
integrals that would arise from considering higher order correlators (like
four-point functions) in the pure GOE plus external channels problem. 

\section{Dotsenko-Fateev Integrals}
The integral in Eq.~(\ref{IMT}) actually bears a close relationship
with a class of integrals introduced first in the context of four-point
correlation functions in 
two-dimensional conformal field theories by Dotsenko \& Fateev 
\cite{DF}. 
The Selberg integral, which is already known to be intimately connected
with the Calogero-Sutherland model and with random-matrix theory in the
guise of the pure GOE problem without coupling to external 
channels \cite{PJF2}, is in fact a special case of a Dotsenko-Fateev (DF) 
integral. 

After appropriate variable transformations chosen to cast it into a 
form suitable for the present application, the general DF integral 
can be written as 
\begin{equation} 
{\cal J}_{nm}(\alpha,\beta;\rho) = \frac{1}{m!n!} 
     \prod_{i=1}^n\int_1^\infty dt_i 
     \prod_{j=1}^m\int_0^1 d\tau_j\, 
     |f_{nm}(\{t_i\},\{\tau_j\};\alpha,\beta,\rho)| \;, 
\label{integral}
\end{equation} 
where 
\begin{equation} 
f_{nm}(\{t_i\},\{\tau_j\};\alpha,\beta,\rho) = 
     \prod_{i=1}^n t_i^{\alpha'} 
     (t_i-1)^{\beta'} \prod_{j=1}^m \tau_j^\alpha(1-\tau_j)^\beta 
     {\cdot} \frac{\prod_{i<i'}(t_i-t_{i'})^{2\rho'} 
     \prod_{j<j'}(\tau_j-\tau_{j'})^{2\rho}}{\prod_{i=1}^n\prod_{j=1}^m 
     (t_i-\tau_j)^2} \;. 
\label{integrand}
\end{equation} 
The constraints on the parameters that ensure solvability of the integral
are given by 
\begin{equation} 
\alpha' = -\rho'\alpha \;,\quad 
     \beta' = -\rho'\beta \;,\quad
     \rho' = \rho^{-1} \;. 
\label{constraints}
\end{equation} 
This is the form that appears in Ref.~\citenum{PJF}, where the result of
its evaluation can be found.
According to Eq.(3.8) of Ref.~\citenum{PJF}, the value of this 
integral can be expressed as
\begin{eqnarray} 
{\cal J}_{nm}(\alpha,\beta;\rho) & = & \rho^{2mn}\prod_{\ell=1}^n 
     \frac{\Gamma(\ell/\rho)}{\Gamma(1/\rho)} \prod_{j=1}^m 
     \frac{\Gamma(j\rho-n)}{\Gamma(\rho)} 
\nonumber \\
& & \cdot \prod_{\ell=0}^{n-1}\frac{\Gamma(1-\beta/\rho+\ell/\rho)
     \Gamma(-1+2m+\alpha/\rho+\beta/\rho-(n-1+\ell)/\rho)}
     {\Gamma(\alpha/\rho-\ell/\rho)} 
\nonumber \\
& & \cdot \prod_{j=0}^{m-1}\frac{\Gamma(1-n+\alpha+j\rho) 
     \Gamma(1-n+\beta+j\rho)}{\Gamma(2-n+\alpha+\beta+(m-1+j)\rho)} \;, 
\label{eval}
\end{eqnarray} 
having implemented the conditions on the parameters 
\mbox{$\alpha',\beta',\rho'$} 
given in Eq.(\ref{constraints}). 

As it turns out, a knowledge of the dependence of the DF integral 
\mbox{${\cal J}_{12}(\alpha,\beta;\rho)$} 
on arbitrary values of its parameters 
\mbox{$\alpha,\beta,\rho$} 
is required in order to evaluate the particular integral expression 
(\ref{IMT}) for $I_M$.
Specifically, we shall need to consider the more general class of
integrals defined by 
\begin{eqnarray} 
J[f] & \equiv & \int_1^\infty dt\, t^{\alpha'}(t-1)^{\beta'} \int_0^1 dt_1\, 
     t_1^\alpha (1-t_1)^\beta \int_0^1 dt_2\, t_2^\alpha (1-t_2)^\beta 
\nonumber \\ 
& & \cdot \frac{|t_1-t_2|^\lambda}{(t-t_1)^2(t-t_2)^2} f(t,t_1,t_2) \;, 
\label{Jf} 
\end{eqnarray}
and we shall require that 
\begin{equation} 
\alpha' = -\frac{2\alpha}{\lambda} \;, \quad
     \beta' = -\frac{2\beta}{\lambda} \;. 
\label{alphabeta}
\end{equation} 
Then, for 
\mbox{$f(t,t_1,t_2) \equiv 1$}, 
the integral 
\mbox{$J[f]$} 
belongs to the class of integrals discussed and evaluated analytically by
Dotsenko \& Fateev, i.e., those encompassed 
by Eqs.~(\ref{integral})--(\ref{constraints}). 
We have 
\mbox{$J[1] = 2{\cal J}_{12}(\alpha,\beta;\rho)$},
and we satisfy the solvability conditions of Eq.~(\ref{constraints}) with the
choice 
\begin{equation} 
\rho' = \frac{2}{\lambda} \;, \quad
     \rho = \frac{\lambda}{2} \;. 
\label{solvability}
\end{equation} 
We note however, that the parameter $\rho$ does not actually enter our
expression because we are considering the special case where there exists
only a single non-compact integration variable (viz.\ $t$). 

What we have encountered in the present application is a moment of the DF
integral. We must choose 
\begin{equation} 
f(t,t_1,t_2) = t-t_1 \;, 
\label{ft12}
\end{equation} 
and 
\begin{eqnarray}
& & \alpha = M/2 \;,\quad \alpha' = -M 
\nonumber \\
& & \beta = -1/2 \;,\quad \beta' = 1 \;. 
\label{parameters} 
\end{eqnarray} 
These latter relations imply that ultimately 
\mbox{$\lambda = 1$} 
must be considered. 
The arbitrariness in $\lambda$ at this stage is
required to ensure the convergence of all intermediate steps. 
Thus, in the framework of Dotsenko \& Fateev, we have 
\mbox{$\rho = 1/2$}, 
\mbox{$\rho' = 2$}, 
which are values characteristic of problems with GOE symmetry.

As mentioned above, one is interested in computing moments of DF
integrals, such as the one defined by Eqs.~(\ref{Jf}) and (\ref{alphabeta}), 
in general since they will arise when considering perturbed GOE's and
making asymptotic expansions of the exact integral result 
in powers of some small parameter. 

\section{The Method of Aomoto} 
We shall approach the problem at hand by method invented by Aomoto 
\cite{Aomoto} 
to solve the Selberg integral. For this purpose, we introduce the function 
\begin{equation} 
D_\lambda(t,t_1,t_2) \equiv t^{\alpha'}(t-1)^{\beta'}(t_1t_2)^\alpha 
     [(1-t_1)(1-t_2)]^\beta\frac{|t_1-t_2|^\lambda} 
     {(t-t_1)^2(t-t_2)^2} \;. 
\label{D12} 
\end{equation} 
Then 
\begin{equation} 
\frac{\partial D_\lambda}{\partial t} = \left[\frac{\alpha'}{t} 
     + \frac{\beta'}{t-1} -\frac{2}{t-t_1} -\frac{2}{t-t_2}
     \right]D_\lambda \;. 
\label{del-t}
\end{equation} 
Using the symmetry of the function
\mbox{$D_\lambda(t,t_1,t_2)$} 
in the variables $t_1$ and $t_2$, and the fact that it vanishes
sufficiently rapidly at the
boundaries of the $t$-integration, viz,
\mbox{$t=1$} 
and $\infty$, we see that 
\begin{eqnarray}
0 & = & \int_1^\infty dt \int_0^1 dt_1 \int_0^1 dt_2\, 
     \frac{\partial}{\partial t}(tD_\lambda) 
\nonumber \\
& = &  \int_1^\infty dt\ \int_0^1 dt_1 \int_0^1 dt_2\,
     \left[ (1+\alpha'+\beta')1 + \beta'\frac{1}{t-1} 
     - 4\frac{t}{t-t_1}\right]D_\lambda \;, 
\label{int-tD}
\end{eqnarray} 
which implies the identity
\begin{equation} 
0 = (1+\alpha'+\beta')J[1] + \beta'J\left[\frac{1}{t-1}\right] 
     -4J\left[\frac{t}{t-t_1}\right] \;. 
\label{(ii)}
\end{equation} 
Likewise, after some algebraic manipulation, we obtain 
\begin{eqnarray} 
0 & = & \int_1^\infty dt \int_0^1 dt_1 \int_0^1 dt_2\, 
     \frac{\partial}{\partial t}(t^2D_\lambda) 
\nonumber \\ 
& = & \int_1^\infty dt \int_0^1 dt_1 \int_0^1 dt_2\,
     \left[ (2+\alpha'+\beta')t + \beta' 1 + \beta'\frac{1}{t-1} 
     - 4\frac{t^2}{t-t_1}\right]D_\lambda \;, 
\label{int-t2D}
\end{eqnarray} 
which leads to a second identity
\begin{equation} 
0 = (2+\alpha'+\beta')J[t] + \beta'J[1] 
     + \beta'J\left[\frac{1}{t-1}\right] 
     - 4J\left[\frac{t^2}{t-t_1}\right] \;. 
\label{(iii)}
\end{equation} 
Now, subtracting Eq.~(\ref{(ii)}) from Eq.~(\ref{(iii)}) serves to
eliminate the terms involving 
\mbox{$J[1/(t-1)]$}, and yields the relationship
\begin{equation} 
0 = (2+\alpha'+\beta')J[t] - (1+\alpha')J[1] 
     +4J\left[\frac{t}{t-t_1}\right] 
     -4J\left[\frac{t^2}{t-t_1}\right] \;. 
\label{(A)}
\end{equation} 

Next, we note that 
\begin{equation} 
\frac{\partial D_\lambda}{\partial t_1} = \left[\frac{\alpha}{t_1} 
     - \frac{\beta}{1-t_1} + \frac{2}{t-t_1} + \frac{\lambda}{t_1-t_2}
     \right]D_\lambda \;. 
\label{del-t1}
\end{equation} 
Thus, upon partial integration, 
\begin{eqnarray}
0 & = & \int_1^\infty dt \int_0^1 dt_1 \int_0^1 dt_2\, 
     \frac{\partial}{\partial t_1}(t_1D_\lambda) 
\nonumber \\
& = & \int_1^\infty dt \int_0^1 dt_1 \int_0^1 dt_2\,
     \left[ (1+\alpha+\beta)1 - \beta\frac{1}{1-t_1} 
     + 2\frac{t_1}{t-t_1} + \frac{\lambda}{2}\right]D_\lambda \;, 
\label{int-t1D}
\end{eqnarray} 
where we have made use of the fact that (owing to the 
\mbox{$t_1 \leftrightarrow t_2$} 
symmetry) 
\begin{eqnarray} 
\int_1^\infty dt \int_0^1 dt_1 \int_0^1 dt_2\, \frac{t_1}{t_1-t_2}D_\lambda 
     & = & \half\int_1^\infty dt \int_0^1 dt_1 \int_0^1 dt_2\, 
     \left[\frac{t_1}{t_1-t_2} + \frac{t_2}{t_2-t_1}\right] D_\lambda
\nonumber \\
& = & \half\int_1^\infty dt \int_0^1 dt_1 \int_0^1 dt_2\, D_\lambda \;. 
\end{eqnarray} 
This yields the identity 
\begin{equation} 
0 = (1+\half\lambda+\alpha+\beta)J[1] - \beta
     J\left[\frac{1}{1-t_1}\right] +2J\left[\frac{t_1}{t-t_1}\right] \;. 
\label{(ii1)}
\end{equation} 
Also, 
\begin{eqnarray}
0 & = & \int_1^\infty dt \int_0^1 dt_1 \int_0^1 dt_2\, 
     \frac{\partial}{\partial t_1}(t_1^2D_\lambda) 
\nonumber \\ 
& = & \int_1^\infty dt \int_0^1 dt_1 \int_0^1 dt_2\,
     \left[ (2+\alpha)t_1 - \beta\frac{t_1^2}{1-t_1} 
     + 2\frac{t_1^2}{t-t_1} + \lambda t_1\right]D_\lambda \;, 
\label{int-t11D}
\end{eqnarray} 
where we have now used the fact that 
\begin{eqnarray} 
\int_1^\infty dt \int_0^1 dt_1 \int_0^1 dt_2\, 
     \frac{t_1^2}{t_1-t_2}D_\lambda 
& = & \half\int_1^\infty dt \int_0^1 dt_1 \int_0^1 dt_2\, \left[
     \frac{t_1^2}{t_1-t_2} + \frac{t_2^2}{t_2-t_1}\right]D_\lambda
\nonumber \\ 
& = & \int_1^\infty dt \int_0^1 dt_1 \int_0^1 dt_2\, t_1 D_\lambda \;.
\end{eqnarray} 
After some further algebraic simplification, we arrive at the identity
\begin{equation} 
0 = (2+\lambda+\alpha+\beta)J[t_1] + \beta J[1] - 
     \beta J\left[\frac{1}{1-t_1}\right] + 
     2J\left[\frac{t_1^2}{t-t_1}\right] \;. 
\label{(iii1)}
\end{equation} 
Subtracting Eq.~(\ref{(ii1)}) from Eq.~(\ref{(iii1)}), we find 
\begin{equation} 
0 = (2+\lambda+\alpha+\beta)J[t_1] - \left(1+\half\lambda+\alpha\right)J[1] -
     2J\left[\frac{t_1}{t-t_1}\right] + 
     2J\left[\frac{t_1^2}{t-t_1}\right] \;, 
\label{(B)} 
\end{equation} 
which again has served to eliminate the terms involving 
\mbox{$J[1/(t-1)]$}. 
To eliminate the terms involving 
\mbox{$J[t_1/(t-t_1)]$}
and 
\mbox{$J[t_1^2/(t-t_1)]$},
let us now consider the addition of Eq.~(\ref{(B)}) to one half times 
Eq.~(\ref{(A)}), which yields the relation 
\begin{equation} 
0 = \left(\half(\alpha'+\beta')-1\right)J[t] + (\lambda+\alpha+\beta)J[t_1] 
     - \left(\half(\lambda-1) + \alpha + \half\alpha'\right)J[1] \;. 
\end{equation} 
It is more convenient to re-express this relationship as 
\begin{equation} 
0 = \left(\half(\alpha'+\beta')-1\right)J[t-1] + 
     (\lambda+\alpha+\beta)J[t_1-1] +
     \left(\half(\lambda-1) + \beta + \half\beta'\right)J[1] \;. 
\end{equation} 

Next, we must set
\mbox{$\alpha' = -2\alpha/\lambda$}, 
\mbox{$\beta' = -2\beta/\lambda$}, 
which leads to 
\begin{eqnarray} 
J[t-1] - \lambda J[t_1-1] & = & \frac{(\lambda-1)(\lambda+2\beta)}
     {2(\alpha+\beta+\lambda)} J[1] 
\nonumber \\ 
& = & \frac{(\lambda-1)^2}{M-1+2\lambda} J[1] \;, 
\label{Jlambda} 
\end{eqnarray} 
having performed the substitutions 
\mbox{$\alpha = M/2$}, 
\mbox{$\beta = -1/2$} 
in the last line. 
Therefore, assuming that the integrals implicit in 
\mbox{$J[t-1]$} 
and 
\mbox{$J[t_1-1]$} 
converge when 
\mbox{$\lambda \to 1^+$}, 
we see that 
\begin{equation} 
J_{\lambda=1}[t-t_1] = \frac{1}{M+1}\lim_{\lambda\to 1^+} 
     (1-\lambda)^2J_\lambda[1] \;. 
\end{equation} 
We can perform the DF integral $J_\lambda[1]$ for $\lambda$ 
in the neighbourhood
of unity by appealing to the results of Ref.~\citenum{PJF}:
If we set 
\mbox{$m=2$}, 
\mbox{$n=1$}, 
\mbox{$\rho=\lambda/2$}, 
then Eq.~(3.8) of Ref.~\citenum{PJF} (or equivalently Eq.~(\ref{eval})
above) implies that 
\begin{equation} 
J_\lambda[1] \asym{\lambda \to 1^+} \frac{4}{(\lambda-1)^2} \;. 
\label{J1} 
\end{equation} 
Accordingly, 
\begin{equation} 
I_M \quad=\quad J_1[t-t_1] \quad=\quad \frac{4}{M+1} \;, 
\end{equation} 
in which case (recalling Eq.~(\ref{IM})), 
\begin{equation} 
\overline{|S_{ab}|^2} = \frac{1}{M+1} 
\end{equation} 
for 
\mbox{$a \neq b$}, 
in agreement with the discussion of Ref.~\citenum{PWZLW}. 
 
It is important to appreciate that because of the singularity in 
$J_\lambda[1]$ at 
\mbox{$\lambda = 1$}, 
the calculation could not have been performed without a knowledge of the
DF integral in a general neighbourhood about 
\mbox{$\lambda =1$}. 
Let us close the discussion with a remark on the physical significance of
the parameters 
\mbox{$\alpha,\beta,\rho$}. 
The value 
\mbox{$\rho = \half$} 
is characteristic of all problems exhibiting an underlying GOE symmetry. 
In pure GOE problems (with no external channels), one typically deals with
\mbox{$\alpha=-1/2, \beta=-1/2$}. 
For example, the density-density correlator 
\mbox{$\overline{\rho(E_1)\rho(E_2)}$}
is given by Eq.~(\ref{Ha-exact}) below with 
\mbox{$x = E_1-E_2$},
\mbox{$t=0$}, 
\mbox{$\lambda=1/2$}, 
\mbox{$p=1$}, 
\mbox{$q=2$} 
and $1/\rho_0$ equal to the mean level spacing. 
After some simple changes of variables, the integrand in
Eq.~(\ref{Ha-exact}) is seen to be of the DF type (\ref{Jf}) with 
\mbox{$\alpha=-1/2, \beta=-1/2$} 
and 
\mbox{$\rho=\half$}. 
Thus we see that the introduction of $M$ external channels promotes
$\alpha$ to a general value. 
It is an open question to determine what sort of additional degrees of
freedom should be incorporated in the random-matrix model in order to 
move $\beta$ away from its default GOE value. 

\chapter{Calogero-Sutherland Model} 
The Calogero-Sutherland model (CSM) describes the quantum mechanics of 
particles on a line interacting through a $1/r^2$ pairwise potential. 
It has been shown that this model is equivalent to a one-dimensional ideal
gas of non-interacting anyons, and its excitations are known to exhibit
fractional statistics. 
For arbitrary rational values of the coupling 
\mbox{$\lambda=p/q$}, 
\mbox{$p,q \in \natnum$}, 
an exact analytical expression has been given by Ha \cite{Ha}, 
in terms of a
\mbox{$(p+q)$}-dimensional 
integral, for the dynamical ground-state
density-density correlation function 
\mbox{$\langle 0|\rho(x,t)\rho(0,0)|0\rangle$}. 
The physical significance of the integers 
\mbox{$p,q$} 
is that an excited state of the CSM involving $p$ elementary
quasi-particle excitations is always accompanied by $q$ quasi-hole
excitations. 
The CSM is closely related to the circular ensembles of random matrix
theory, with the ground-state wavefunctions of the CSM reproducing the
eigenvalue distributions of random matrix ensembles corresponding to
orthogonal, unitary and symplectic symmetry at couplings of 
\mbox{$\lambda = 1/2, 1, 2$}, 
respectively. 
Also, after a Wick rotation in the time variable $t$, the dynamical 
density-density correlation at these couplings coincides with the dynamical
density-density correlation for the parameter-dependent GOE, GUE and
GSE, respectively \cite{SA}. 
It is not too surprising then, to discover that the DF integral has a role
to play in the calculation of physical quantities in the CSM.

Starting with the general form of the DF integral, we shall consider a
limiting form obtained as the result of a certain scaling. 
This will lead us to an exact identity which allows one to directly
compute the value of the
multi-dimensional integral that represents the overall coefficient in the
asymptotic behaviour of Ha's density-density correlation function in the
limit of large spatial and temporal separations. 

\section{An Integral Identity}
Suppose that in the DF integral, as given by Eqs.~(\ref{integral}) and 
(\ref{integrand}), 
we set 
\mbox{$t_i = \exp\{s_i/\alpha\}$} 
and 
\linebreak 
\mbox{$\tau_j = \exp\{-\xi_j/\alpha\}$}
after eliminating the parameters 
\mbox{$\alpha',\beta',\rho'$}
with the aid of Eq.~(\ref{constraints}),  
and then consider the limit 
\mbox{$\alpha\to\infty$}. 
In this limit, we can write
\pagebreak 
\begin{eqnarray} 
f_{nm}(\{e^{s_i/\rho}\},\{e^{-\xi_j/\rho}\};\alpha,\beta,\rho)
     & {\displaystyle \asym{\alpha\to\infty}} &
     \prod_{i=1}^n (s_i/\alpha)^{-\beta/\rho} 
     e^{-s_i/\rho} \prod_{j=1}^m (\xi_j/\alpha)^\beta e^{-\xi_j} 
\nonumber \\
& &  \cdot 
     \frac{\prod_{i<i'}\alpha^{-2/\rho}(s_i-s_{i'})^{2/\rho} 
     \prod_{j<j'}\alpha^{-2\rho}(\xi_{j'}-\xi_j)^{2\rho}} 
     {\prod_{i=1}^n\prod_{j=1}^m \alpha^{-2mn}(s_i+\xi_j)^2} \;. 
\end{eqnarray} 
Noting also that 
\begin{equation} 
\int_1^\infty dt_i \asym{\alpha\to\infty} \frac{1}{\alpha}\int_0^\infty 
     ds_i \;,\quad 
     \int_0^1 d\tau_j \asym{\alpha\to\infty} \frac{1}{\alpha} 
     \int_0^\infty d\xi_j \;, 
\end{equation} 
we see that Eq.~(\ref{integral}) becomes 
\begin{eqnarray} 
{\cal J}_{nm}(\alpha,\beta;\rho) & {\displaystyle \asym{\alpha\to\infty}} &
     \alpha^{-c_{nm}(\beta,\rho)} \prod_{i=1}^n\int_0^\infty ds_i\, 
     s_i^{-\beta/\rho}e^{-s_i/\rho} \prod_{j=1}^m \int_0^\infty d\xi_j\, 
     \xi_j^\beta e^{-\xi_j} 
\nonumber \\
& &  \cdot 
     \frac{\left|\prod_{i<i'}(s_i-s_{i'})^{2/\rho} \prod_{j<j'} 
     (\xi_{j'}-\xi_j)^{2\rho}\right|}
     {\prod_{i=1}^n\prod_{j=1}^m (s_i+\xi_j)^2}\;, 
\label{integral2}
\end{eqnarray} 
where 
\begin{equation} 
c_{nm}(\beta,\rho) \equiv m+n+m\beta-n\beta/\rho+m(m-1)\rho+n(n-1)/\rho 
     -2mn \;. 
\end{equation} 

On the other hand, using the relation
\begin{equation}
\frac{\Gamma(a+x)}{\Gamma(x)} \asym{x \to\infty} x^a \;, 
\end{equation} 
we see that 
\begin{eqnarray} 
\frac{\Gamma(1-n+\alpha+j\rho)}{\Gamma(2-n+\alpha+\beta+(m-1+j)\rho)} 
     & {\displaystyle \asym{\alpha\to\infty}} & \frac{\Gamma(\alpha)}
     {\Gamma(\alpha+1+\beta+(m-1)\rho)} 
\nonumber \\ 
& {\displaystyle \asym{\alpha\to\infty}} & \frac{1}{\alpha^{1+\beta 
     +(m-1)\rho}} \;, 
\end{eqnarray} 
and 
\begin{eqnarray} 
\frac{\Gamma(-1+2m+\alpha/\rho+\beta/\rho-(n-1+\ell)/\rho)}
     {\Gamma(\alpha/\rho-\ell/\rho)} 
     & {\displaystyle \asym{\alpha\to\infty}} & 
     \frac{\Gamma(\alpha/\rho-1+2m+\beta/\rho-(n-1)/\rho)} 
     {\Gamma(\alpha/\rho)} 
\nonumber \\
& {\displaystyle \asym{\alpha\to\infty}} & 
     \left(\frac{\rho}{\alpha}\right)^{1-2m-\beta/\rho+(n-1)/\rho} \;. 
\end{eqnarray} 
It follows from Eq.~(\ref{eval}) that 
\begin{eqnarray} 
{\cal J}_{nm}(\alpha,\beta;\rho) & {\displaystyle \asym{\alpha\to\infty}} &
     \alpha^{-c_{nm}(\beta,\rho)} \rho^{(1-\beta/\rho+(n-1)/\rho)n} 
     \prod_{\ell=1}^n\frac{\Gamma(\ell/\rho)}{\Gamma(1/\rho)} 
     \prod_{j=1}^m\frac{\Gamma(j\rho-n)}{\Gamma(\rho)} 
\nonumber \\ 
& & \cdot \prod_{\ell=0}^{n-1}\Gamma(1-\beta/\rho+\ell/\rho) 
     \prod_{j=0}^{m-1}\Gamma(1-n+\beta+j\rho) \;. 
\label{eval2}
\end{eqnarray} 
After equating Eq.~(\ref{integral2}) with Eq.~(\ref{eval2}) in the limit 
\mbox{$\alpha\to\infty$}, 
we finally obtain the integral identity 
\begin{eqnarray} 
I_{nm}(z;\beta,\rho) & \equiv & \frac{1}{m!n!}
     \prod_{i=1}^n\int_0^\infty dr_i\, r_i^{-\beta/\rho} e^{-zr_i} 
     \prod_{j=1}^m\int_0^\infty d\eta_j\, \eta_j^\beta e^{-z\eta_j} 
     \frac{\left|\prod_{i<i'}(r_i-r_{i'})^{2/\rho}
     \prod_{j<j'}(\eta_{j'}-\eta_j)^{2\rho}\right|}
     {\prod_{i=1}^n\prod_{j=1}^m (\eta_j+\rho r_i)^2} 
\nonumber \\ 
& = & z^{-c_{nm}(\beta,\rho)} 
     \prod_{\ell=1}^n\frac{\Gamma(\ell/\rho)}{\Gamma(1/\rho)} 
     \prod_{j=1}^m\frac{\Gamma(j\rho-n)}{\Gamma(\rho)} 
     {\cdot} \prod_{\ell=0}^{n-1}\Gamma(1-\beta/\rho+\ell/\rho) 
     \prod_{j=0}^{m-1}\Gamma(1-n+\beta+j\rho) \;,
\nonumber \\ & &  
\label{identity}
\end{eqnarray} 
having also introduced the common scaling factor $z$ and used it in making
the change of integration variables
\mbox{$s_i/\rho = zr_i$}, 
\mbox{$\xi_j = z\eta_j$}. 

\section{Ha's Result}
The exact result of Ha \cite{Ha} for the dynamical ground-state 
density-density correlator reads
\begin{eqnarray} 
\langle 0|\rho(x,t)\rho(0,0)|0\rangle  & = & C\prod_{i=1}^q\int_0^\infty 
     dx_i \prod_{j=1}^p \int_0^1 dy_j\, Q^2 \frac{\left|\prod_{i<i'} 
     (x_i-x_{i'})^{2\lambda} \prod_{j<j'} (y_j-y_{j'})^{2/\lambda}\right|} 
     {\prod_{i=1}^q \prod_{j=1}^p(x_i+\lambda y_j)^2} 
\nonumber \\ 
& & \cdot \prod_{i=1}^q\left[x_i(x_i+\lambda)\right]^{\lambda-1} 
     \prod_{j=1}^p\left[\lambda y_j(1-y_j)\right]^{1/\lambda-1} 
     \cos Qx \exp \{-iEt\} \;, 
\label{Ha-exact}
\end{eqnarray} 
where 
\begin{eqnarray} 
Q & = & 2\pi\rho_0\biggl(\sum_{i=1}^q x_i + \sum_{j=1}^p y_j \biggr) \;, 
\nonumber \\ 
E & = & (2\pi\rho_0)^2\biggl(\sum_{i=1}^q x_i(x_i+\lambda) 
     + \sum_{j=1}^p \lambda y_j(1-y_j) \biggr) \;, 
\nonumber \\ 
C & = & \frac{\lambda^{2p(q-1)}\Gamma^2(p)}{2\pi^2p!q!} 
     \frac{\Gamma^q(\lambda)\Gamma^p(1/\lambda)} 
     {\prod_{i=1}^q\Gamma^2(p-\lambda(i-1)) \prod_{j=1}^p 
     \Gamma^2(1-(j-1)/\lambda)} \;, 
\label{QEC}
\end{eqnarray} 
and 
\mbox{$p,q$} are related by 
\mbox{$p/q = \lambda$}. 
One should note that, in the formulae given above, the symbol $\rho_0$
denotes the average density of states for the system 
\mbox{$\rho_0 = N/L$} ($N$ being the total number of particles and $L$ the
one-dimensional volume of the system), and should not be
confused with the parameter of similar name $\rho$ in the DF integral.  

In the region of large $x$ and $t$, the integral above is saturated by
neighbourhoods of all points satisfying
\mbox{$x_i = 0$} 
for all $i$ and 
\mbox{$y_j = \mbox{0 or 1}$}. 
However, any point with 
\mbox{$y_j = 1$} 
for some 
\mbox{$j = 1,2,\ldots,p$} 
will give rise to an oscillating contribution. Thus, 
in order to calculate the leading-order non-oscillatory term in the limit
of large 
\mbox{$x,t$}, 
we need to expand all the variables in the integrand about the end-points
\mbox{$x_i,y_j = 0$}. 
This yields 
\begin{eqnarray} 
\langle 0|\rho(x,t)\rho(0,0)|0\rangle & {\displaystyle \asym{x,t\to\infty}} & 
%
%
     \half (2\pi\rho_0)^2C{\cal A}(q,p,\lambda)
     \left[\frac{1}{(2\pi i\rho_0 x -i (2\pi\rho_0)^2\lambda t)^2} + 
     \frac{1}{(2\pi i\rho_0 x +i (2\pi\rho_0)^2\lambda t)^2}\right] \;, 
\nonumber \\ & & 
\label{asym}
\end{eqnarray} 
where 
\begin{eqnarray} 
{\cal A}(q,p,\lambda) & \equiv & 
     \prod_{i=1}^q \int_0^\infty dx_i\, x_i^{\lambda-1}e^{-x_i} 
     \prod_{j=1}^p \int_0^\infty dy_j\, y_j^{1/\lambda-1}e^{-y_j} 
     \left(\frac{Q}{2\pi\rho_0}\right)^2 
\nonumber \\
& &  \cdot \frac{\left|
     \prod_{i<i'}(x_i-x_{i'})^{2\lambda}\prod_{j<j'}(y_j-y_{j'})^{2/\lambda} 
     \right|}
     {\prod_{i=1}^q\prod_{j=1}^p (x_i+\lambda y_j)^2} \;. 
\label{APQ}
\end{eqnarray} 
In carrying out the final steps in deriving Eq.~(\ref{asym}), 
the $y_j$ integration intervals have been extended to infinity and then 
changes of integration variables have
been made in order to remove the $x$ and $t$ dependent factors from the
exponentials. Additionally, since the exponentials were initially
imaginary, a Wick rotation is also assumed to have been performed without
obstruction from significant singularities in the complex $x_i$ and $y_j$
planes. The fact the 
\mbox{$\lambda = p/q$} 
has been used too. 

\section{Asymptotic Limit}
The task now is to compute the multiple integral 
\mbox{${\cal A}(q,p,\lambda)$} 
of Eq.~(\ref{APQ}). 
It is clear that differentiating the integral identity of 
Eq.~(\ref{identity}) twice with respect to $z$ and
subsequently setting 
\mbox{$z=1$} 
will yield an integral of the required type. 
We must choose 
\mbox{$\beta=\rho-1$}, 
\mbox{$m=q$}, 
\mbox{$n=p$} 
and 
\mbox{$\rho=(p+\epsilon)/q$}, 
with the limit 
\mbox{$\epsilon \to 0$} 
to be taken later. 
Then, 
\begin{equation} 
{\cal A}(q,p,\lambda=p/q) = p!q!\left.\frac{\partial^2}{\partial z^2} 
     I_{pq}(z;\rho-1,\rho)\right|_{z=1} \;, 
\label{deriv}
\end{equation} 
having let 
\mbox{$\epsilon \to 0$} 
on the RHS. 
Now, to leading order in $\epsilon$, we find
\begin{equation} 
z^{-c_{pq}(\beta,\rho)} 
     \asym{\epsilon\to 0} z^{-\epsilon^2(q/p)} \;. 
\end{equation} 
Thus, 
\begin{equation} 
I_{pq}(z;\rho-1,\rho) \asym{\epsilon\to 0} 
     z^{-\epsilon^2(q/p)}\frac{\Gamma^2(\epsilon)} 
     {\left[\Gamma(p/q)\right]^q} \prod_{\ell=1}^p 
     \frac{\Gamma^2(q\ell/p)}{\Gamma(q/p)} \prod_{j=1}^{q-1} 
     \Gamma^2(jp/q-p) \;, 
\label{epsilon} 
\end{equation} 
recalling that 
\mbox{$\rho=(p+\epsilon)/q$}. 
After differentiating twice with respect to $z$, setting 
\mbox{$z=1$} 
and taking the limit 
\mbox{$\epsilon\to 0$}, 
we obtain the evaluation of the integral which occurs in the asymptotic
expansion, i.e., 
\begin{equation} 
{\cal A}(q,p,p/q) = \frac{q!p!}{\left[\Gamma(p/q)\right]^q} 
     \left(\frac{q}{p}\right) \prod_{\ell=1}^p 
     \frac{\Gamma^2(q\ell/p)}{\Gamma(q/p)} \prod_{j=1}^{q-1} 
     \Gamma^2(jp/q-p) \;. 
\label{result}
\end{equation} 
We note that the integral identity (\ref{identity}) could not be used 
directly with the final desired choice of parameters 
\mbox{$\rho = \lambda = p/q$} 
because, as seen from Eq.~(\ref{epsilon}), this case is singular. 
Only after the twofold differentiation can the final parameter values be
set. Hence, as in the previous calculation, a knowledge of the DF-integral
based identity (\ref{identity}) 
is required in an arbitrary neighbourhood of the relevant
point in the parameter space. 

Let us now express the normalization constant $C$ in Eq.~(\ref{QEC}) 
in a form similar to the product of gamma functions appearing above. 
We have 
\begin{eqnarray} 
\frac{1}{\prod_{i=1}^q\Gamma^2(p-\lambda(i-1))} & = & 
     \frac{\lambda^{-2p(q-1)}}{\Gamma^2(p)} 
     {\cdot} \frac{1}{\prod_{i=2}^q\Gamma^2(-\lambda(i-1))} 
\nonumber \\
& & \cdot \frac{1}{\Bigl[\prod_{i=2}^q\left((p-1)/\lambda -(i-1)\right) 
     \left((p-2)/\lambda -(i-1)\right)\cdots \left(-(i-1)\right)\Bigr]^2} 
\nonumber \\ 
& = & \frac{\lambda^{-2p(q-1)}}{\Gamma^2(p)\Gamma^2(q)} 
     {\cdot} \frac{1}{\prod_{i=2}^q\Gamma^2(-\lambda(i-1))}
     \prod_{j=2}^p\left[\frac{\Gamma((j-1)/\lambda-(q-1))} 
     {\Gamma((j-1)/\lambda)}\right]^2 \;. 
\end{eqnarray} 
But, 
\begin{eqnarray} 
\prod_{j=2}^p \Gamma^2((j-1)/\lambda-(q-1)) & = & \prod_{j=1}^p 
     \Gamma^2(1-(j-1)/\lambda) \;, 
\nonumber \\ 
\frac{1}{\Gamma^2(q)} \prod_{j=2}^p \frac{1}{\Gamma^2((j-1)/\lambda)} 
     & = & \prod_{j=1}^p\frac{1}{\Gamma^2(j/\lambda)} \;, 
\end{eqnarray} 
and 
\begin{equation} 
\prod_{i=2}^q \Gamma^2(-\lambda(i-1)) = \prod_{i=1}^{q-1} 
     \Gamma^2(i\lambda-p) \;, 
\end{equation} 
having made extensive use of the relation 
\mbox{$\lambda = p/q$}. 
Therefore, 
\begin{equation} 
\frac{1}{\prod_{i=1}^q \Gamma^2(p-\lambda(i-1))} = 
     \frac{\lambda^{-2p(q-1)}}{\Gamma^2(p)} {\cdot}
     \frac{1}{\prod_{i=1}^{q-1}\Gamma^2(i\lambda-p)}
     \prod_{j=1}^p \frac{\Gamma^2(1-(j-1)/\lambda)}{\Gamma^2(j/\lambda)} \;.
\end{equation} 
Hence, 
\begin{equation} 
C = \frac{1}{2\pi^2p!q!}\frac{\Gamma^q(\lambda)\Gamma^p(1/\lambda)} 
     {\prod_{j=1}^p\Gamma^2(j/\lambda) 
     \prod_{i=1}^{q-1}\Gamma^2(i\lambda-p)} \;. 
\end{equation} 

Substitution of this formula for $C$ into the asymptotic expansion above
(i.e.\ Eq.~(\ref{asym})), 
together with the evaluation of the multiple integral 
\mbox{${\cal A}(q,p,\lambda)$} 
in Eq.~(\ref{result}), gives 
\begin{equation} 
\langle 0|\rho(x,t)\rho(0,0)|0\rangle \asym{x,t \to\infty} 
     -\frac{\rho_0^2}{\lambda}\left[\frac{1}{(2\pi\rho_0 x - 
     (2\pi\rho_0)^2\lambda t)^2} + \frac{1} 
     {(2\pi\rho_0 x + (2\pi\rho_0)^2\lambda t)^2}\right] \;, 
\end{equation} 
which is the required result, i.e.\ Eq.~(7.10) of Ref.~\citenum{Ha}. 
In that paper, the relationship between the normalization constant $C$ and
the integral 
\mbox{${\cal A}(q,p,\lambda)$}, which are multiplied together in 
Eq.~(\ref{asym}) to yield the simple factor 
\mbox{$C{\cal A} = (2\pi^2\lambda)^{-1}$}, 
had to be deduced by indirect means. 
Here, with the aid of the DF integral, we have constructed an explicit 
derivation of this result. 

\section*{Acknowledgements} 
J.A.Z.\ wishes to thank the members of the Mathematics Department at the
University of Melbourne for their hospitality during visits when most of
this work was done. J.A.Z. also acknowledges financial support from NSERC
(Canada) in the earlier stages of this project.


\begin{thebibliography}{99}
\bibitem{Meh91}
M.L.~Mehta, {\it Random Matrices}, 2{\it nd} edition, (Academic Press, 
     New York, 1991).
\bibitem{VWZ} 
J.J.M.~Verbaarschot, H.A.~Weidenm\"uller \& M.R.~Zirnbauer, 
     Phys.~Rep. {\bf 129}~(1985)~367.
\bibitem{PWZL} 
Z.~Pluha\v{r}, H.A.~Weidenm\"uller, J.A.~Zuk \& C.H.~Lewenkopf, 
     Phys.~Rev.~Lett. {\bf 73}~(1994)~2115. 
\bibitem{IWZ} 
S.~Iida, H.A.~Weidenm\"uller \& J.A.~Zuk, 
     Ann.~Phys. {\bf 200}~(1990)~219.
\bibitem{BW} 
O.~Bohigas \& H.A.~Weidenm\"uller, Ann.~Rev.~Nucl.~Part.~Sci. 
     {\bf 38}~(1988)~421. 
\bibitem{LSF} 
P.A.~Lee, A.D.~Stone \& H.~Fukuyama, Phys.~Rev. {\bf B35}~(1987)~1039.
\bibitem{HAW}
H.A.~Weidenm\"uller, Nucl.~Phys. {\bf A522}~(1991)~293c.
\bibitem{Efetov} 
K.B.~Efetov, Adv.~Phys. {\bf 32}~(1983)~53. 
\bibitem{PWZLW} 
Z.~Pluha\v{r}, H.A.~Weidenm\"uller, J.A.~Zuk, C.H.~Lewenkopf \& 
F.J.~Wegner, Ann.~Phys. {\bf 243}~(1995)~1.
\bibitem{DF}
V.S.~Dotsenko \& V.A.~Fateev, Nucl.~Phys. {\bf B251}~(1985)~691.
\bibitem{PJF2} 
P.J.~Forrester, Nucl.~Phys. {\bf B388}~(1992)~671. 
\bibitem{PJF} 
P.J.~Forrester, Mod.~Phys.~Lett. {\bf B9}~(1995)~359.
\bibitem{Aomoto}
K.~Aomoto, SIAM~J.~Math.~Analysis {\bf 18}~(1987)~545. 
\bibitem{Ha} 
Z.N.C.~Ha, Nucl.~Phys. {\bf B435}~(1995)~604.
\bibitem{SA} 
B.D.~Simons \& B.L.~Altshuler, Phys.~Rev. {\bf B48}~(1993)~5422.
\end{thebibliography}
\end{document}